\documentclass[conference]{IEEEtran}
\IEEEoverridecommandlockouts
\usepackage{amsmath,amssymb,amsfonts}
\usepackage{algorithmic}
\usepackage{graphicx}
\usepackage{textcomp}
\usepackage{xcolor}
\usepackage[hidelinks]{hyperref}
\usepackage{siunitx}
\usepackage{lipsum}
\usepackage{booktabs}
\usepackage{array}    
\usepackage{amssymb}
\usepackage{caption}
\usepackage[acronym]{glossaries}
\usepackage[a4paper, total={184mm,239mm}]{geometry}
\usepackage{cite}
\usepackage{placeins}

\def\BibTeX{{\rm B\kern-.05em{\sc i\kern-.025em b}\kern-.08em
\kern-.1667em\lower.7ex\hbox{E}\kern-.125emX}}

\newcommand{\figref}[1]{Fig.~\ref{#1}}
\DeclareSIUnit{\cycle}{cycle}

\begin{document}

\title{Distributed Inference with Minimal Off-Chip Traffic for Transformers on Low-Power MCUs}

\author{
  \IEEEauthorblockN{
    Severin Bochem\IEEEauthorrefmark{1}, 
    Victor J.B. Jung\IEEEauthorrefmark{2}, 
    Arpan Suravi Prasad\IEEEauthorrefmark{2}, 
    Francesco Conti\IEEEauthorrefmark{3}, 
    Luca Benini\IEEEauthorrefmark{2}\IEEEauthorrefmark{3}
  }
  \IEEEauthorblockA{
    \IEEEauthorrefmark{1}\textit{D-ITET, ETH Zurich}, Switzerland;
    \IEEEauthorrefmark{2}\textit{Integrated Systems Laboratory, ETH Zurich}, Switzerland; \\
    \IEEEauthorrefmark{3}\textit{DEI, and Information Engineering, University of Bologna}, Italy 
  }
  \IEEEauthorblockA{
    sbochem@ethz.ch, \{jungvi, prasadar, lbenini\}@iis.ee.ethz.ch, f.conti@unibo.it
  }
}

\maketitle
\newacronym{AR}{AR}{Augmented Reality}
\newacronym{VR}{VR}{Virtual Reality}
\newacronym{XR}{XR}{Extended Reality}
\newacronym{GEMM}{GEMM}{General Matrix Multiply}
\newacronym{GEMV}{GEMV}{General Matrix-Vector Multiply}
\newacronym{EDP}{EDP}{Energy Delay Product}
\newacronym{MIPI}{MIPI}{Mobile Industry Processor Interface}
\newacronym{SPMD}{SPMD}{Single Program Multiple Data}
\newacronym{FM}{FM}{Foundation Model}
\newacronym{AI}{AI}{Artificial Intelligence}
\newacronym{DL}{DL}{Deep Learning}
\newacronym{DNN}{DNN}{Deep Neural Network}
\newacronym[longplural={Application-Specific Integrated Circuits}]{ASIC}{ASIC}{Application-Specific Integrated Circuit}
\newacronym[longplural={Graphics Processing Units}]{GPU}{GPU}{Graphics Processing Unit}
\newacronym{TPU}{TPU}{Tensor Processing Unit}
\newacronym{CPU}{CPU}{Central Processing Unit}
\newacronym[longplural={Instruction Set Architectures}]{ISA}{ISA}{Instruction Set Architecture}
\newacronym[plural=TCNs, firstplural={Temporal Convolutional Neural Networks (TCNs)}]{TCN}{TCN}{Temporal Convolutional Neural Network}
\newacronym[plural=CNNs, firstplural={Convolutional Neural Networks (CNNs)}]{CNN}{CNN}{Convolutional Neural Network}
\newacronym{MHSA}{MHSA}{Multi-Head Self-Attention}
\newacronym{MHA}{MHA}{Multi-Head Attention}
\newacronym[longplural={Reduced Instruction Set Computers}]{RISC}{RISC}{Reduced Instruction Set Computer}
\newacronym{ARM}{ARM}{Advanced RISC Machine}
\newacronym{KV}{KV}{Key-Value}
\newacronym{SSR}{SSR}{Stream Semantic Register}
\newacronym{DSP}{DSP}{Digital Signal Processing}
\newacronym{HW}{HW}{Hardware}
\newacronym{WL}{WL}{Workload}
\newacronym{ONNX}{ONNX}{Open Neural Network Exchange}
\newacronym{NAS}{NAS}{Neural Architecture Search}
\newacronym{PULP}{PULP}{Parallel Ultra Low Power}
\newacronym{AXI}{AXI}{Advanced eXtensible Interface}
\newacronym{TCDM}{TCDM}{Tightly Coupled Data Memory}
\newacronym[longplural={Micro-Controller Units}]{MCU}{MCU}{Micro-Controller Unit}
\newacronym{ODL}{ODL}{On-Device Learning}
\newacronym{SGD}{SGD}{Stochastic Gradient Descent}
\newacronym{FLOPS}{FLOPS}{Floating Point Operations Per Second}
\newacronym{SNN}{SNN}{Spiking Neural Network}
\newacronym{KWS}{KWS}{Keyword Spotting}
\newacronym{IIS}{IIS}{Integrated Systems Laboratory}
\newacronym[longplural={Large Language Models}]{LLM}{LLM}{Large Language Model}
\newacronym[longplural={Systems-on-Chip}]{SoC}{SoC}{System-on-Chip}
\newacronym{NLP}{NLP}{Natural Language Processing}
\newacronym{SotA}{SotA}{State of the Art}
\newacronym{CV}{CV}{Computer Vision}
\newacronym{MAC}{MAC}{Multiply-Accumulate}
\newacronym{IoT}{IoT}{Internet of Things}
\newacronym{SIMD}{SIMD}{Single Instruction Multiple Data}
\newacronym{PTQ}{PTQ}{Post-Training Quantization}
\newacronym{QAT}{QAT}{Quantization Aware Training}
\newacronym{EEG}{EEG}{Electroencephalogram}
\newacronym{RAW}{RAW}{Read-After-Write}
\newacronym{WRL}{WRL}{Weight-Reuse Linear}
\newacronym{IRL}{IRL}{Input-Reuse Linear}
\newacronym{LWT}{LWT}{Layer-Wise Tiling}
\newacronym{DFT}{DFT}{Depth-First Tiling}
\newacronym{MQA}{MQA}{Multi-Query Attention}
\newacronym{GQA}{GQA}{Grouped-Query Attention}
\newacronym[longplural={Vision Transformers}]{ViT}{ViT}{Vision Transformer}
\newacronym{GELU}{GELU}{Gaussian Error Linear Unit}
\newacronym{HPC}{HPC}{High-Performance Computing}
\newacronym[longplural={Direct Memory Accesses}]{DMA}{DMA}{Direct Memory Access}
\newacronym{PE}{PE}{Processing Element}
\newacronym{FPU}{FPU}{Floating-Point Unit}
\newacronym{RAM}{RAM}{Random-Access Memory}
\newacronym{SRAM}{SRAM}{Static Random-Access Memory}
\newacronym{NE16}{NE16}{Neural Engine 16-channels}
\newacronym{FWSA}{FWSA}{Fused-Weight Self-Attention}
\newacronym{TQT}{TQT}{Trained Quantization Threshold}
\newacronym{ReLU}{ReLU}{Rectified Linear Unit}
\newacronym{ML}{ML}{Machine Learning}
\newacronym{SLM}{SLM}{Small Language Model}
\newacronym{FC}{FC}{Fully-Connected}
\newacronym{MLP}{MLP}{Multi-Layer Perceptron}
\newacronym{AXI}{AXI}{Advanced eXtensible Interface}
\begin{abstract}
Contextual \gls{AI} based on emerging Transformer models is predicted to drive the next technology revolution in interactive wearable devices such as new-generation smart glasses. By coupling numerous sensors with small, low-power 
\glspl{MCU}, these devices will enable on-device intelligence and sensor control. A major bottleneck in this class of systems is the small amount of on-chip memory available in the \glspl{MCU}. In this paper, we propose a methodology to deploy real-world Transformers on low-power wearable devices with minimal off-chip traffic exploiting a distributed system of \glspl{MCU}, partitioning inference across multiple devices and enabling execution with stationary on-chip weights. We validate the scheme by deploying the TinyLlama-42M decoder-only model on a system of 8 parallel ultra-low-power \glspl{MCU}. The distributed system achieves an energy consumption of 0.64\,mJ, a latency of 0.54\,ms per inference, a super-linear speedup of 26.1\,$\times$, and an \gls{EDP} improvement of 27.2\,$\times$, compared to a single-chip system. On MobileBERT, the distributed system's runtime is 38.8\,ms, with a super-linear 4.7\,$\times$ speedup when using 4 \glspl{MCU} compared to a single-chip system.
\end{abstract}

\section{Introduction}
Transformer models \cite{vaswani2023attentionneed} have revolutionized the landscape of \gls{AI} by achieving breakthroughs in areas such as \gls{NLP} and \gls{CV} \cite{xie2021segformersimpleefficientdesign}. The success of transformer-based language models such as BERT \cite{devlin2019bertpretrainingdeepbidirectional}, GPT \cite{openai2024gpt4technicalreport}, or Llama \cite{touvron2023llamaopenefficientfoundation} is largely due to their capability to capture contextual relationships within data, which makes them particularly appealing to use in contextual \gls{AI} tasks commonly found in smart glasses, including personalized assistance and context-aware interactions. 

Despite their success, deploying these Transformers on resource-constrained devices at the extreme edge presents formidable challenges, resulting from their high computational and memory requirements. Conventional Transformer models, which feature many millions to many billions of parameters \cite{devlin2019bertpretrainingdeepbidirectional, ramesh2022hierarchicaltextconditionalimagegeneration}, are inherently too large to fit within the computation and memory budget of edge devices, necessitating reliance on off-chip memory or even cloud services. This dependency results in higher latency, increased power consumption, and privacy concerns, all critical in wearable devices.
Smart glasses represent a promising wearable platform with the potential to enhance user experience through contextual \gls{AI} \cite{konrad2024gazegptaugmentinghumancapabilities}. By enabling seamless interaction with the environment, they could provide users with context-aware responses that enhance everyday life. However, deploying \gls{LLM}s on such edge devices is infeasible due to their size and computational requirements. Tackling this challenge, \gls{SLM}s with tens to a few hundred million, rather than several billion parameters have been proposed \cite{eldan2023tinystoriessmalllanguagemodels, zhang2024tinyllamaopensourcesmalllanguage, sun2020mobilebertcompacttaskagnosticbert}. Still, even for \gls{SLM}s, one main bottleneck during Transformer inference on smart glasses is the limitation in on-chip memory, which typically does not exceed 8 MiB \cite{kwon2023xrbenchextendedrealityxr}. Even for small Transformer models, weights and intermediate tensors might need to be stored and accessed from off-chip memory, which is both latency and energy-intensive. 

Previous works have explored the distribution of Transformer models on multiple compute units, thus leveraging vast computational resources to execute intensive \gls{LLM} workloads. These distributed methods \cite{aminabadi2022deepspeedinferenceenablingefficient, pope2022efficientlyscalingtransformerinference} allow Transformer models to meet their computational and memory requirements by partitioning workloads across multiple nodes, thereby overcoming the limitations of a single compute node. 
However, most of these works target high-performance computer architectures like  \glspl{CPU}, \glspl{GPU} or \glspl{TPU}.
While data centers focus on increasing throughput and parallelism by batching tokens from multiple users to reduce the memory boundedness of the workload, the same methods cannot be applied to edge devices that require real-time and sequential processing. 
Moreover, edge devices such as smart glasses are subject to very strict constraints on latency, power, and form factor compared to cloud systems. 
This discrepancy calls for a novel approach to deploy Transformers at the edge, requiring careful optimization of latency, power consumption, and form-factor constraints for devices like smart glasses.

To tackle the challenges above, we propose a distributed inference scheme to facilitate the efficient deployment of \glspl{SLM} on resource-constrained \glspl{SoC}.

We deploy our scheme on a network of Siracusa \cite{prasad2024siracusa16nmheterogenous} chips designed for smart glasses, featuring a cluster of 8 parallel RISC-V cores with instruction extensions for \gls{ML} and \gls{DSP} workloads.

Our approach enables running TinyLlama with 42 million parameters~\cite{karpathy2023llama2c} and MobileBERT~\cite{sun2020mobilebertcompacttaskagnosticbert} models solely from on-chip memory \cite{groq}, with minimal overhead associated with inter-chip communication. The main contributions of our paper include:

\begin{itemize}
    \item A strategy to partition the Transformers' Decoder and Encoder onto a distributed system of \glspl{MCU}. This scheme minimizes chip-to-chip communication and needs only two synchronizations per Transformer block. The weights are scattered and never duplicated to reduce the on-chip memory footprint. This strategy enables individual Transformer blocks to be run only from on-chip memory, leading to lower energy per inference and super-linear latency reduction.
    \item Benchmarking of our partitioning strategy for the autoregressive and prompt mode of the decoder-only TinyLlama model as well as MobileBERT's encoder. We extended our results with a scalability study on up to 64 Siracusa \glspl{MCU} to test the limits of our approach.
\end{itemize}

We perform experiments using the open-source event-driven simulator GVSoC \cite{GVSoC}. From the simulator, we extract latencies and the number of accesses to different memory levels, which are fed into an analytical model to estimate the system energy.
Our partitioning improves the performance of autoregressive Tinyllama inference by $26.1 \times$, while incurring a similar energy per inference when using $8$ chips compared to a single chip. This demonstrates a super-linear scaling for the autoregressive TinyLlama mode, as it relies solely on on-chip memory to run a single Transformer layer. By eliminating long-latency off-chip memory accesses during inference, we achieve the aforementioned super-linear speedup.
A scaled-up model achieves $60.1 \times$ performance improvement and $1.3 \times$ energy reduction for $64$ chips, showing our approach's scalability to larger networks. 
In the prompt mode of TinyLlama, using $8$ chips improves performances more than linearly by $9.9 \times$. Finally, for the MobileBERT model, using $4$ chips improves performance by $4.7 \times$ per chip without costing any additional energy per inference.

\section{Background}\label{sec:background}

\subsection{Transformer Networks}
Transformers have revolutionized the field of \gls{NLP} and achieved \gls{SotA} performance in many other domains, such as \gls{CV}. In \gls{NLP}, encoder-decoder and decoder-only models dominate, while for \gls{CV}, mostly encoder-only models are employed. Despite their large memory and compute demands, Transformers have found use in resource-constrained environments \cite{MCU_transformer}. 
The two main building blocks of a Transformer are the \gls{MHSA} and the Full-Connected Layer. Due to its computing intensity and high memory footprint, the \gls{MHSA} is the most challenging to deploy, especially on resource-constraint devices.

The dimensions specifying the Transformer operations are the \textit{sequence length S}, the \textit{embedding dimension E}, the \textit{projection dimension P}, and the \textit{head dimension H}. The first step projects the input $X \in \mathbb{R}^{S \times E}$ onto the queries, keys, and values, $Q, K, V \in \mathbb{R}^{S \times P}$ as shown in equation \ref{eq:attention_mechanism}.
\begin{equation}
\mathbf{Q}=\mathbf{X} \mathbf{W}_{\text {query }}, \quad \mathbf{K}=\mathbf{X W}_{\text {key }}, \quad \mathbf{V}=\mathbf{X} \mathbf{W}_{\text {value }}
\label{eq:attention_mechanism}
\end{equation}
In the Attention step, $Q$, $K$ and $V$ are combined by
\begin{equation}
\operatorname{Attention}(\mathbf{Q}, \mathbf{K}, \mathbf{V}):=\operatorname{softmax}\left(\frac{\mathbf{Q K}^{\mathrm{T}}}{\sqrt{d}}\right) \mathbf{V}, 
\end{equation}
where $d$ is the dimension of $K$ used to scale the Attention. The softmax function is applied to each row of the matrix and defined as 
\begin{equation}
\operatorname{softmax}(\boldsymbol{x})_i=\frac{e^{x_i-\max (\boldsymbol{x})}}{\sum_{j=1}^n e^{x_j-\max (\boldsymbol{x})}}
\end{equation}
for the \textit{i-th} element of a row of size $n$. This operation is performed independently for each of the $H$ heads. 

The output of the \gls{MHSA} is fed into the Fully-Connected Layer consisting of two Linear layers, a row-wise normalization, and a \gls{GELU} \cite{hendrycks2023gaussianerrorlinearunits}. The shapes of the weight matrices in the linear layers are $E \times F$ and $F \times E$ respectively, where $F$ is the \textit{intermediate} dimension of the Transformer model.

This paper focuses on two different modes of Transformer inference, namely autoregressive and prompt mode. 
In autoregressive mode, each output token is predicted sequentially, based on all the previously predicted tokens using a data structure called \gls{KV}-Cache to store results of previous computations.
In prompt mode, multiple outputs get predicted from multiple inputs simultaneously in one inference. Therefore, the main kernel of prompt mode inference is a \gls{GEMM}, whereas in autoregressive mode \gls{GEMV} operations are dominant, which implies that prompt mode is more computationally intensive than autoregressive mode.

\begin{table*}[t]
\centering
\footnotesize 
\renewcommand{\arraystretch}{1.2}
\caption{Comparison of \gls{SotA} works on model partitioning of machine learning inference}
\label{tab:comparison}
\begin{tabular}{@{}cccccc@{}} 
\toprule 
Work & Model & Scale & Platform & Pipelining & Weight Duplication \\
\midrule 
Deepthings \cite{deepthings} & CNN & Low-Power & Raspberry Pi & No & Yes \\
Efficiently Scaling Transformer Inference \cite{pope2022efficientlyscalingtransformerinference} & Transformer & Datacenter & TPU & No & No \\
DeepSpeed Inference \cite{aminabadi2022deepspeedinferenceenablingefficient} & Transformer & Datacenter & GPU & Yes & No \\
When the Edge Meets Transformers \cite{edge_transformer} & Transformer & Low-Power & CPU & No & Yes \\
Hermes \cite{han2024hermesmemoryefficientpipelineinference} & Transformer & Low-Power & CPU & Yes & No \\
Ours & Transformer & Extreme Edge & Siracusa (MCU) & No & No \\
\bottomrule 
\end{tabular}
\end{table*}

\subsection{Deployment platform}\label{deployment_platform}
\begin{figure}
    \centering
    \includegraphics[width=\columnwidth]{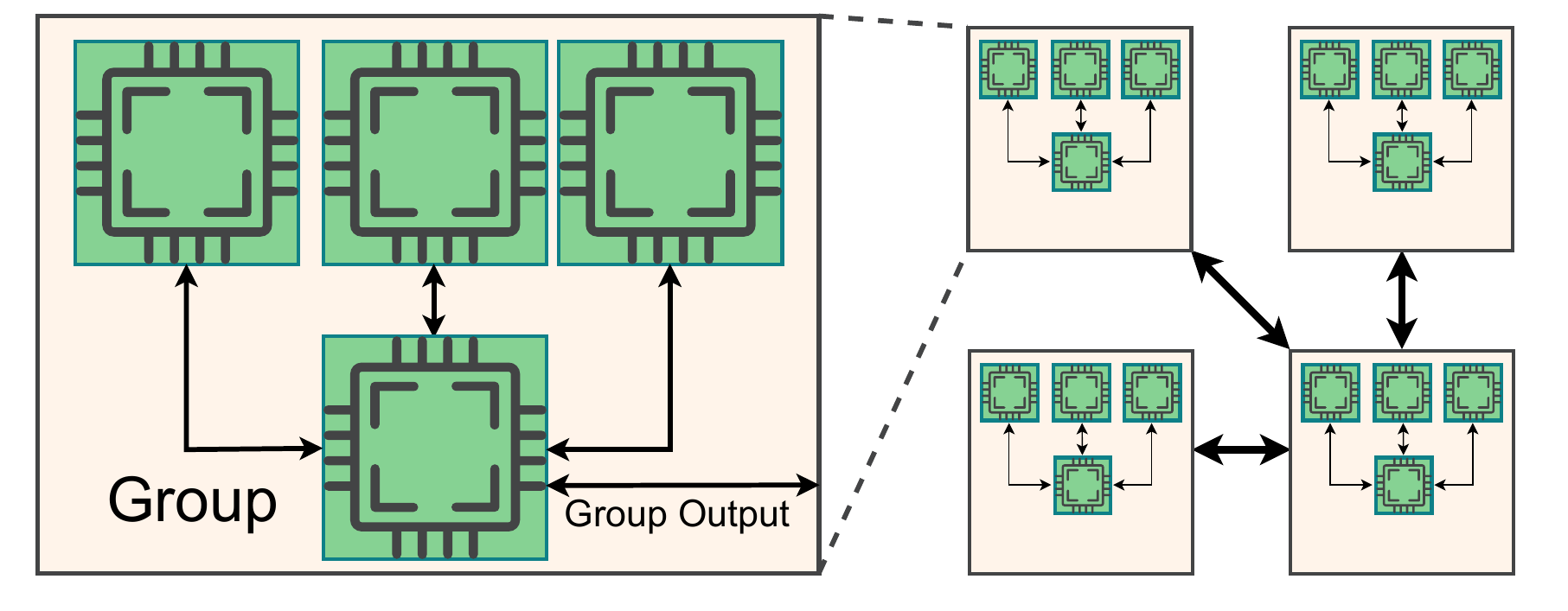}
    \caption{Hierarchical interconnection of the Siracusa chips in the proposed system. Chips are placed in groups of four for improved scalability of the system. We use \gls{MIPI} for the chip-to-chip link.}
    \label{fig:Multi-chip}
\end{figure}

We partition the model on a multi-chip architecture consisting of multiple generic Siracusa chips as shown in \figref{fig:Multi-chip}. As chip-to-chip link, we use the \gls{MIPI} serial interface with $100$ \si{\pico\joule\per\byte} energy consumption and $0.5$ \si{\giga\byte\per\second} bandwidth. All-reduce operations are performed hierarchically in groups of four to reduce the contention on the interconnect, as shown in \figref{fig:Multi-chip}.

\begin{figure}
    \centering
    \includegraphics[width=0.9\linewidth]{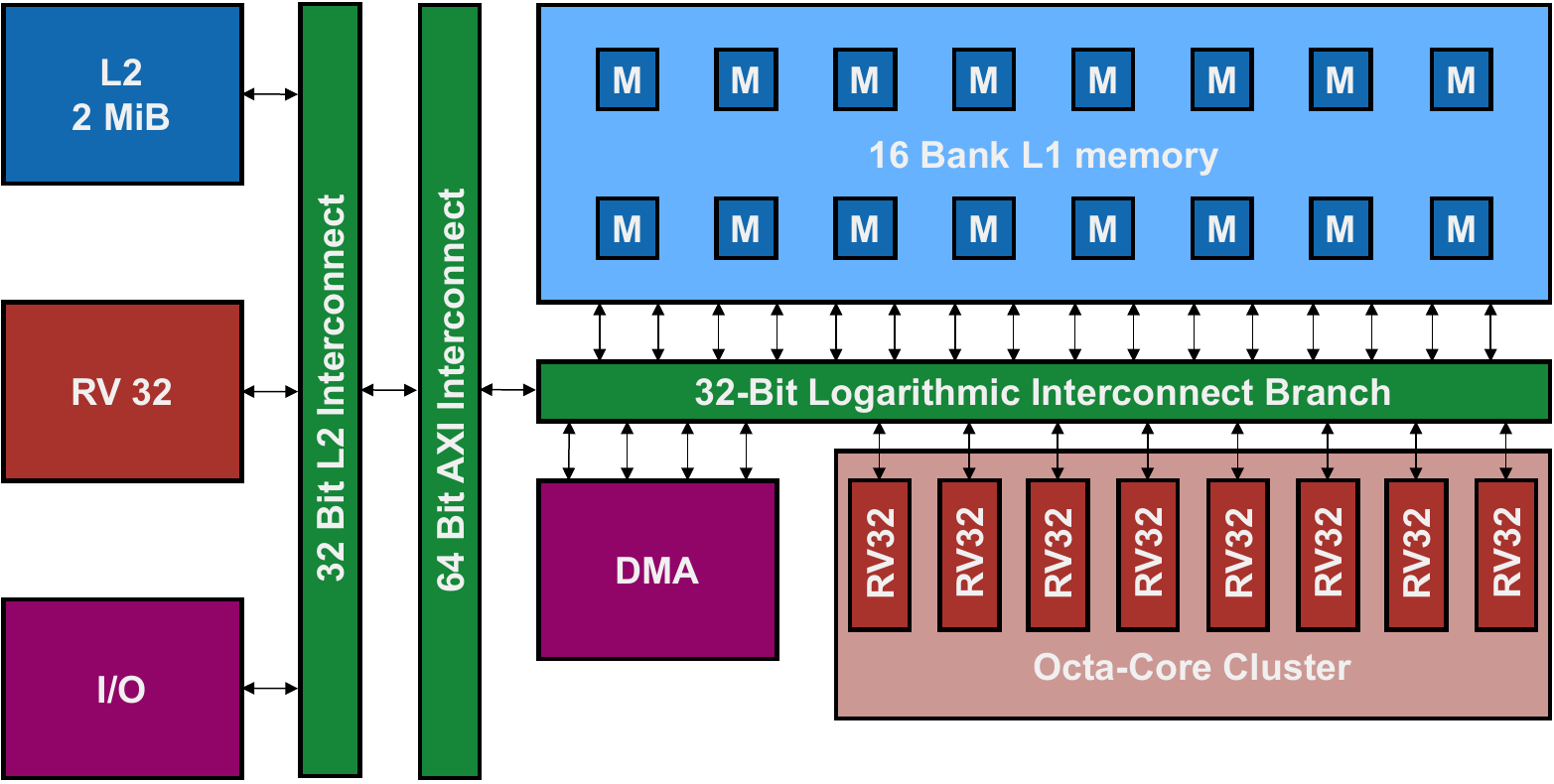}
    \caption{Overview of the generic Siracusa \gls{SoC} including an octa-core RISC-V cluster and host controller (red), memory hierarchy with two levels of scratchpad memory, two arbitrated interconnects towards the L1 memory and an \gls{AXI} interconnect (green), and peripherals such as the cluster \gls{DMA} and chip-level I/O (purple). Note that the image does not depict the N-EUREKA accelerator, as it was not used in this work.}
    \label{fig:siracusa_arch}
\end{figure}

Each chip of the multi-chip architecture is a Siracusa \cite{prasad2024siracusa16nmheterogenous} low-power, heterogeneous RISC-V \gls{MCU} which features an accelerator cluster of eight RISC-V cores, enabling \gls{SPMD} processing~\cite{prasad2024siracusa16nmheterogenous, GAP-8}. An overview of the Siracusa architecture is depicted in \figref{fig:siracusa_arch}. To keep assumptions about the deployment platform minimal and the setup general, we do not use Siracusa's N-EUREKA accelerator. 
To enable single-latency access from cluster cores to the L1 \gls{TCDM}, the cores are connected to the 16 L1 memory banks through a logarithmic interconnect using one 32-bit port each, granting a total memory bandwidth of \SI{256}{bit\per\cycle} to the compute cluster.
While each chip is equipped with significant computing capabilities, its on-chip memory is not sufficient to run inference of \glspl{SLM} such as MobileBERT~\cite{sun2020mobilebertcompacttaskagnosticbert} or TinyLlama~\cite{zhang2024tinyllamaopensourcesmalllanguage}.

\section{Related Work} \label{sec:related_work}

\subsection{Small Language Models}
\glspl{FM}, such as decoder-only \glspl{LLM}, like Llama \cite{touvron2023llama2openfoundation} and Mixtral \cite{jiang2024mixtralexperts} come with large compute and memory demands, often requiring TBs of storage which makes them challenging to deploy on edge devices. 
\glspl{SLM} address this gap by condensing Large Language Models \glspl{LLM} into tens to hundreds of MBs. Some notable examples of \glspl{SLM} include TinyLlama \cite{zhang2024tinyllamaopensourcesmalllanguage}, the Phi series \cite{li2023textbooksneediiphi15, gunasekar2023textbooksneed} and MobileLLM \cite{liu2024mobilellmoptimizingsubbillionparameter}. Methods like incorporating high-quality data \cite{gunasekar2023textbooksneed} and structured pruning techniques \cite{xia2024shearedllamaacceleratinglanguage} aim to improve the efficacy of \glspl{SLM}.

Embedding \glspl{FM} into edge devices may enable a new wave of intelligent,
responsive, and autonomous devices such as smart glasses. 
This work contributes to the goal of efficiently deploying \glspl{SLM} on edge devices by proposing and benchmarking a partitioning scheme that can be applied to a wide range of FMs, from autoregressive decoder-only to
encoder-only ones.

\subsection{Distributed Model Inference}

\begin{figure*}[h!]
    \centering
    \includegraphics[page=1, width=\linewidth]{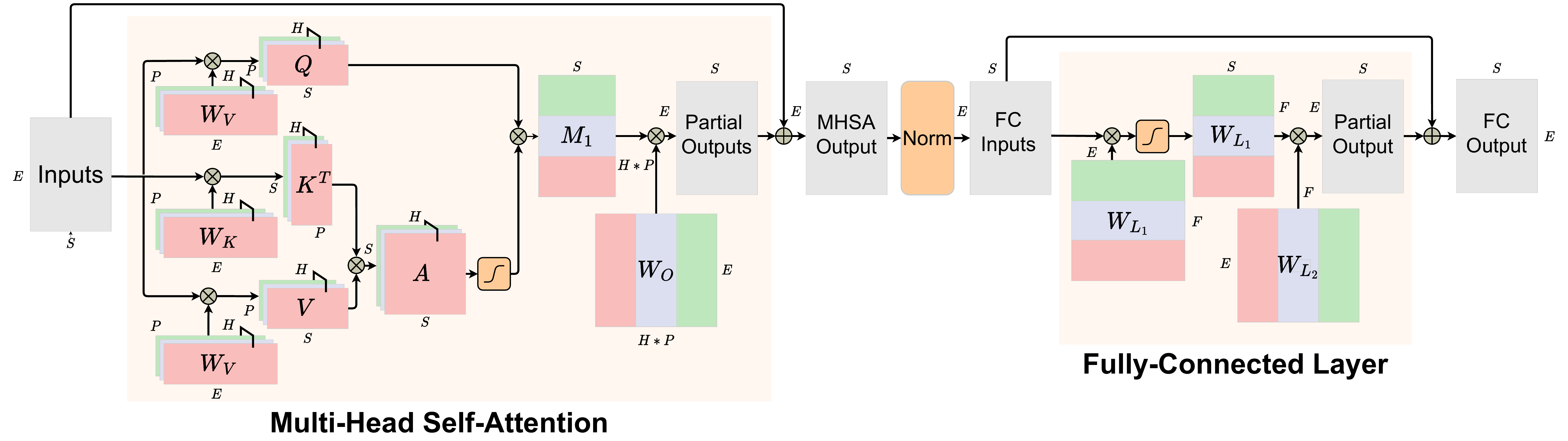}
    \caption{Partitioning of Transformer Inference for three chips. Tensor colorings indicate on which chip a tensor is present. Tensors with grey coloring are present in all chips. Softmax and Norm are shown in orange.}
    \label{fig:MHSA_partitioning}
\end{figure*}

One main bottleneck of Transformer inference on edge devices is the available on-chip memory that can be used to store model weights and intermediate tensors. Each Siracusa chip used in this work features only \SI{256}{\kibi\byte}  in L1 and \SI{2}{\mebi\byte} in L2 memory (See \ref{deployment_platform}).

For \glspl{DNN}, a commonly used method to overcome the bottleneck of available on-chip memory is to partition the inference workload across multiple devices to reduce the memory and compute demands for each chip. Deepthings partition \gls{CNN} inference across multiple \gls{IoT} devices by splitting its input feature maps  \cite{deepthings}. Follow-up works like EdgeFlow\cite{hao2021edgeflowachievingpracticalinteractive} introduced support for network and device heterogeneity. However, these methods are all tailored towards \gls{CNN} inference and are not directly applicable to Transformer models. 
\cite{edge_transformer}.
Recent work from Google partitions Transformer inference across multiple \glspl{TPU} \cite{pope2022efficientlyscalingtransformerinference} tailored towards data center applications and inference of models with more than 500 Billion parameters. This makes memory considerations vastly different from the inference of small models at the edge.
Groq proposes a software-defined datacenter-scale system that aims to minimize off-chip memory access during inference \cite{groq}.
PipeEdge \cite{PipeEdge} partitions Transformer models on edge devices leveraging pipeline parallelism. Hermes\cite{han2024hermesmemoryefficientpipelineinference} chooses a similar pipeline parallel approach. However, pipeline parallelism is infeasible for real-time single-user applications like smart glasses as it requires a sufficient batch size to keep the pipeline utilized and is unable to optimize the latency of an individual request.
Another work \cite{edge_transformer} that aims for low-power Transformer inference targets \gls{CPU} applications and needs to replicate model weights across devices. While this approach can reduce computational demands, the reliance on off-chip memory persists.
An overview of previous works on distributed model inference can be found in table \ref{tab:comparison}.

In this work, we propose a tensor parallelism-based distributed inference scheme across Siracusa chips to facilitate the efficient deployment of \glspl{SLM} on resource-constraint low-power \glspl{MCU}. By not having to replicate any model weights, this partitioning enables running TinyLlama \cite{zhang2024tinyllamaopensourcesmalllanguage} and MobileBERT \cite{sun2020mobilebertcompacttaskagnosticbert} solely from on-chip memory. This is especially beneficial in models that are bound by memory rather than compute latency, such as the autoregressive mode of TinyLlama. With previous approaches, these models would not fit into on-chip memory \cite{edge_transformer} or would lead to an insufficient chip utilization \cite{PipeEdge} for real-time inference. This partitioning scheme does not face the high communication cost common for tensor parallelism, as communication between chips is minimized.

\section{Partitioning Scheme} \label{sec:methods}
A visualization of our partitioning of the \gls{MHSA} can be found in \ref{fig:MHSA_partitioning}. In this example, we assume an \gls{MHSA} with $3$ attention heads distributed across $3$ chips for visualization purposes. The input to the \gls{MHSA} is broadcast to all chips. The weight matrices $W_Q$, $W_K$, and $W_V$ are evenly split across chips, which results in each chip holding one slice of the weight tensors of shape $E\times P \times \frac{H}{Num\_Chips}$, divided across the attention head dimension. Note that in \figref{fig:MHSA_partitioning}, we assume $H = Num\_Chips = 3$ for ease of visualization. Each chip will hold a partition of dimension $S\times P \times \frac{H}{Num\_Chips}$ of the tensors $Q$, $K$ and $V$. Partitioning the \gls{MHSA} across the head dimension is favorable, as the computations along $H$ are fully independent of one another, requiring the chips to communicate only once after the \gls{MHSA}. 

Each chip holds a slice of the $W_O$ matrix of shape $\frac{H*P}{Num\_Chips} \times E$, which is applied to a slice of the intermediate tensor of shape $S \times \frac{H*P}{Num\_Chips}$. After the partial \gls{MHSA}, each chip holds a partial output of shape $S \times E$, which means that an all-reduce operation is needed before the normalization can be applied. As an all-to-one reduce operation lacks the required scalability, we perform a hierarchical reduction in groups of chips. First, a reduction is applied in a group of four chips by sending all partial outputs to one specific chip of the group, on which the partial outputs are accumulated. The outputs of this reduction are then again reduced until the final output of the \gls{MHSA} is computed on one of the chips as visualized in \figref{fig:Multi-chip}. The skip connection from the \gls{MHSA} input to the output shown in \figref{fig:MHSA_partitioning} can be merged into the all-reduce operation as all chips hold the full input tensor. After this output is normalized on a single chip, it is then broadcast back to all chips in the same manner as it is reduced. 

For the \gls{FC} layer, we perform a similar approach. Both weight matrices of the fully connected stage $W_{L_1}$ and $W_{L_2}$ are sliced across the $F$ dimension across chips, requiring no weight replication and resulting in each chip holding a slice of shape $E \times \frac{F}{Num\_Chips}$ of the $W_{L_1}$ tensor and a slice of shape $\frac{F}{Num\_Chips} \times E$ of the $W_{L_2}$ tensor. Similar to the \gls{MHSA}, each chip produces a partial output of shape $S \times E$. From these partial outputs, the final output is produced in an all-reduce operation while also considering the skip connection. Note that this partitioning scheme replicates no weights across chips, which is crucial to save in on-chip memory of edge devices for Transformer applications. Furthermore, it requires only two synchronizations of chips at the end of the \gls{MHSA} and fully connected layer. 

\begin{figure*}[h!]
    \centering
    \includegraphics[page=1, width=\linewidth]{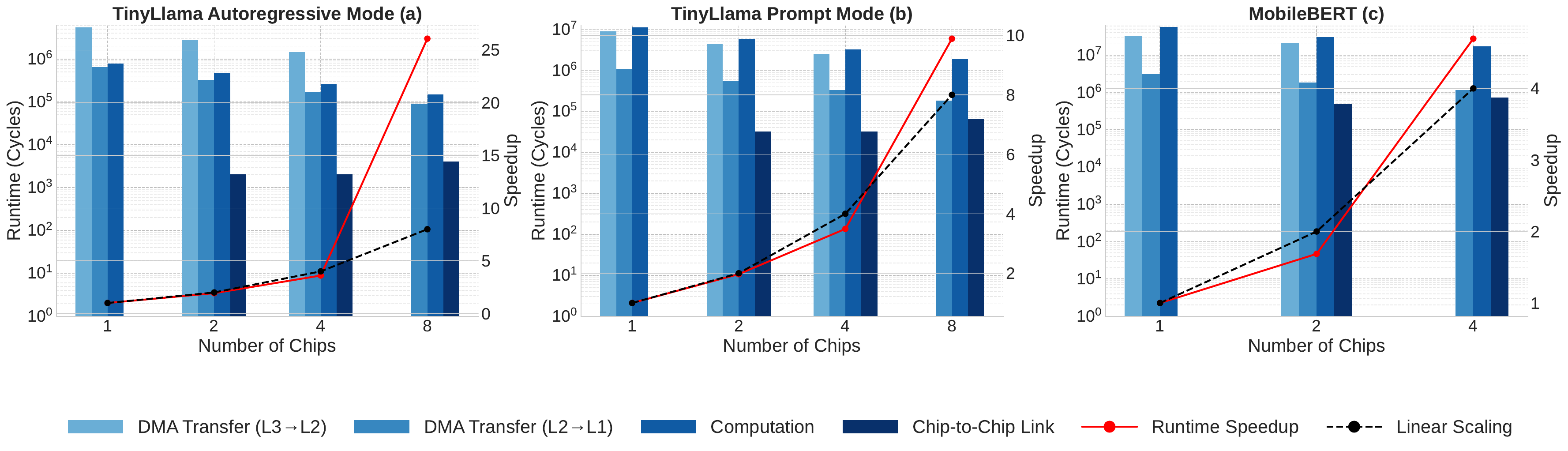}
    \caption{Results of the MobileBERT model and TinyLlama model in prompt and autoregressive modes. The lines indicate the speedup when using $1$-$8$ for TinyLlama or $1$-$4$ for MobileBERT compared to a single-chip system. The bar plot shows the breakdown of runtime into computation, chip-to-chip communication, and access to L2 and L3 memory.}
    \label{fig:plots_all}
\end{figure*}

\section{Evaluation \& Results} \label{sec:eval_results}

\subsection{Experimental Setup} \label{sec:setup}

We conduct experiments using the open-source event-driven simulator GVSoC \cite{GVSoC} to emulate the multi-chip architecture consisting of multiple Siracusa-like chips.
From GVSoC, we obtain the latency and the number of accesses to each memory level, which is used by an analytical model to extract the energy consumption.
For chip-to-chip interconnects, we use an analytical model of \gls{MIPI} with $100$ \si{\pico\joule\per\byte} energy consumption and $0.5$ \si{\giga\byte\per\second} bandwidth \cite{gomez2022distributedonsensorcomputearvr}.
The energy is computed analytically, assuming $100$ \si{\pico\joule\per\byte} for accessing L3 memory and $2$ \si{\pico\joule\per\byte} for accessing L2 memory.
The average power consumption of one core is $13$\,\si{\milli\watt}~\cite{prasad2024siracusa16nmheterogenous} and the cluster of each \gls{SoC} runs at $500$ \si{\mega\hertz} \cite{prasad2024siracusa16nmheterogenous}. The total system energy is computed as follows:

$$
\begin{aligned}
    E_{\text {Total }}= & N_{C2C}*E_{2C2} + \sum_{j=1}^{\#(\text { Chips })} P*T_{Comp,j}\\
    & +N_{L_3 \leftrightarrow L_2,j}*E_{L_3 \leftrightarrow L_2,j} +N_{L_2 \leftrightarrow L_1,j}*E_{L_2 \leftrightarrow L_1,j}.
\end{aligned}
$$ 

Where $P$ is the average power consumption, $T_{Comp,j}$ denotes the computation time of the chip $j$, $N_{C2C}$ is the number of chip-to-chip transfer, $N_{L_2 \leftrightarrow L_1,j}$ and $E_{L_2 \leftrightarrow L_1,j}$ are the number of transfers and the transfer energy between L1 and L2 for the chip $j$, respectively.

To deploy the model partitions, we extend the open-source \gls{ONNX} compiler Deeploy~\cite{scherer2024deeployenablingenergyefficientdeployment} that is tailored for Transformer models on edge devices. As workloads, we run a TinyLlama~\cite{karpathy2023llama2c} and MobileBERT~\cite{sun2020mobilebertcompacttaskagnosticbert} model. 
We take the TinyLlama model from an open-source implementation with an embedding dimension $E$ of $512$, an intermediate size of $2048$, and $8$ layers, matching the configuration of the model released initially. In autoregressive mode, this model leverages a KV-Cache to avoid unnecessary recomputation.
We distribute the TinyLlama model across up to $8$ chips. For our scalability study, we use a modified version of TinyLlama, containing $64$ heads, and perform inference distributed on up to $64$ chips. To do so, we leave all other model parameters unchanged. 
We use TinyLlama with a sequence length of $128$ for autoregressive mode and $16$ for prompt mode.
The MobileBERT model has an embedding dimension and intermediate size of $512$, $4$ attention heads, and a sequence length of $268$. 

In our experiments, we depict the runtime and energy for a single Transformer block.
The weights of the next Transformer block are loaded into L2 memory from L3 memory during the execution of the current block in a double-buffered fashion.

\begin{figure*}[h!]
    \centering
    \includegraphics[page=1, width=\linewidth]{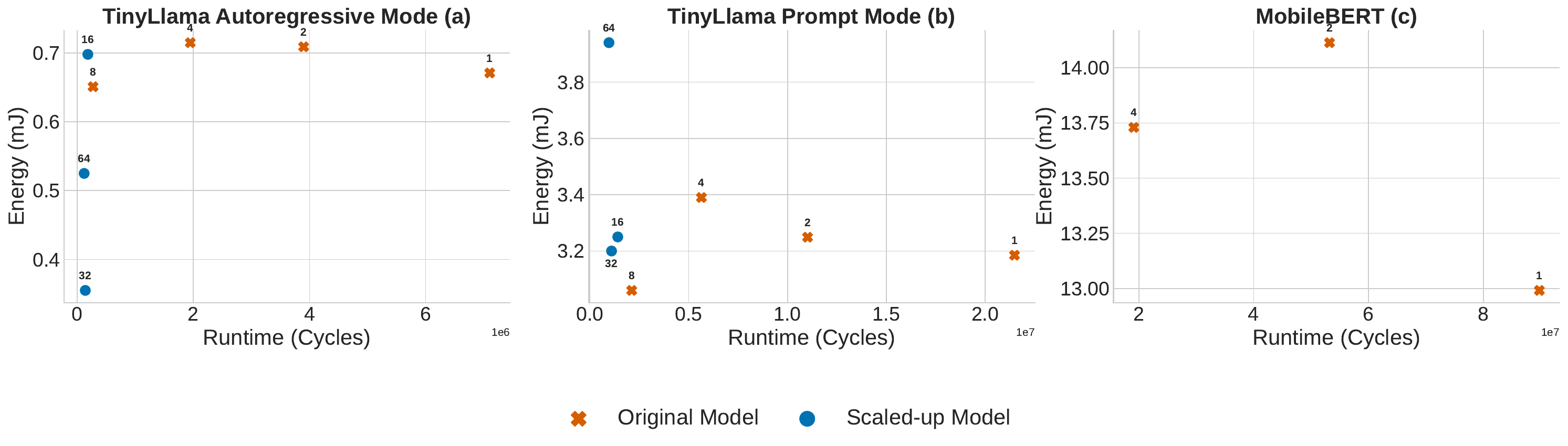}
    \caption{This figure depicts runtimes and energies of TinyLlama in autoregressive mode (left), TinyLlama in prompt mode (middle), and MobileBERT (right). Red crosses are results obtained for models in their default configuration, whereas red circles show results for the scaled-up models.}
    \label{fig:plots_energy_latency}
\end{figure*}

\subsection{Runtime and Energy Consumption}

In the following subsection, we showcase the results of our partitioning scheme with the setup and networks described in \ref{sec:setup}. 
First, we partition the autoregressive and prompt modes of TinyLlama and MobileBERT models in their original configuration across Siracusa chips. \figref{fig:plots_all} shows runtime speedup results and a runtime breakdown for all three models. 
\figref{fig:plots_energy_latency} depicts the energy and latency for all three models in a 2D plot. Note that \figref{fig:plots_energy_latency} also contains results of our scalability study that we address in Sec. \ref{subsec:scalability}.

In autoregressive mode, we achieve a speedup of $26.1 \times$, when using $8$ chips, compared to using only a single chip, resulting in a super-linear scaling as seen in \figref{fig:plots_all} (a).
Super-linear speedup is not achieved for 1, 2, and 4 chips because the model weights of one TinyLlama block are too large to fit on the aggregated on-chip memory.
Hence, for 1, 2, and 4 chips, many off-chip transfers are required in the execution of one transformer block, and they are the major contributor to the total runtime.
\figref{fig:plots_energy_latency} (a) shows that using $8$ chips reduces the energy consumption per inference.
This is a consequence of minimizing the chip-to-chip connection, not replicating model weights across chips, and storing intermediate tensors in L2 instead of L3. 

In prompt mode, inference latency is reduced by $9.9 \times$ when using $8$ chips over a single chip, which again leads to a super-linear runtime scaling as shown in \figref{fig:plots_all} (b).
\figref{fig:plots_energy_latency} (b) shows that the energy consumption is reduced when using $8$ chips, as, similar to autoregressive mode, we don't need off-chip transfer to process the current layer when $8$ or more chips are used.
\figref{fig:plots_all} (a) and (b) show clearly that in autoregressive mode, accessing memory is the main contributor to overall runtime, whereas, in prompt mode, computation is the largest contributor.
Therefore, in prompt mode, reducing the number of off-chip transfers to L3 leads to less speedup compared to autoregressive mode, as the workload is not bottlenecked by off-chip memory transfer in the first place.

Finally, \figref{fig:plots_all} (c) and \figref{fig:plots_energy_latency} (c) depict the latency and energy for the partitioning of MobileBERT.
Partitioning on $4$ chips results in a super-linear speedup of $4.7 \times$ due to the suppression of off-chip transfers to L3. 
However, using $4$ chips results in a slight increase in inference energy. 
This is caused by the increased partitioning that scales down the kernel size of the Transformer. Therefore, it becomes more challenging to achieve high utilization of the RISC-V cores in each chip, which slightly hurts energy efficiency. In particular, for example, the runtime of a \gls{GEMM} kernel does not scale down linearly as the overall kernel size is reduced, resulting in a runtime reduction that is less than linear at the network scale. 

\begin{figure}[t]
    \centering
    \includegraphics[page=1, width=0.8\linewidth]{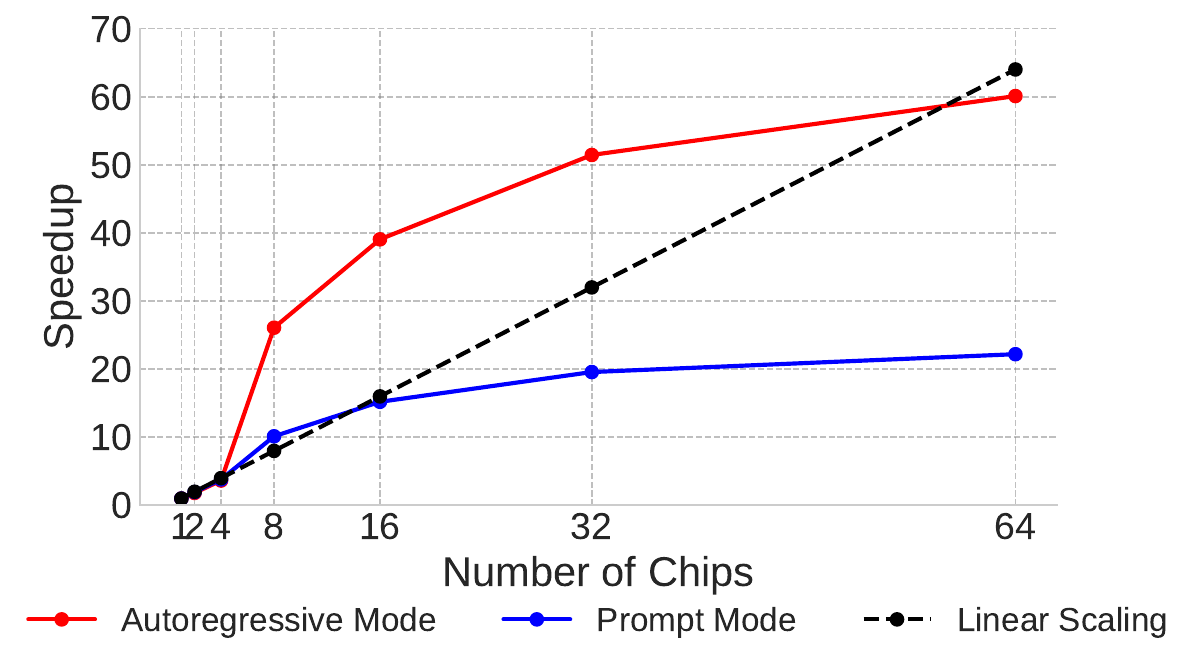}
    \caption{Speedup on a scaled up TinyLlama model on $2$-$64$ chips compared to a single-chip system.}
    \label{fig:combined_speedup_scaled}
\end{figure}

\subsection{Scalability Study} \label{subsec:scalability}

Next, we study the scalability of our partitioning scheme to a larger number of chips. We increase the number of heads of the TinyLlama model from $8$ to $64$ while keeping the other parameters constant. 
\figref{fig:combined_speedup_scaled} shows the speedup of the scaled-up model for both the autoregressive and prompt mode up to 64 chips.

In autoregressive mode, we achieve a speedup of $60.1 \times$ using $64$ chips instead of a single-chip system, demonstrating that our partitioning scheme achieves a quasi-linear speedup. Additionally, the energy consumption per inference is reduced by $1.3 \times$ as shown in \figref{fig:plots_all} (a).
For $8$ and $16$ chips, we achieve a super-linear speedup for an individual Transformer block as one block can be run accessing only on-chip memory, whereas, for $1$, $2$, and $4$ chips, off-chip memory is required to hold model weights and intermediate tensors of the current block. 
During the processing of one layer, the weights of the next layer can be loaded into on-chip memory, incurring an additional energy penalty. 
However, with $32$ chips, all model weights fit on-chip, and double-buffering is no longer required, resulting in a further energy reduction that can be observed in \figref{fig:plots_energy_latency} (a). 

In the prompt mode of TinyLlama inference, we achieve a linear speedup up until a $16$-chip system as seen in \figref{fig:combined_speedup_scaled}.
Scaling the system further has diminished returns as the prompt mode is dominated by computation, and saving in off-chip memory accesses has a reduced benefit compared to autoregressive mode. 
Furthermore, the \gls{GEMM} kernel's runtime scale is sub-linearly as the dimensions are reduced. Additionally, the number of chip-to-chip transfers and the accumulation of partial tensors introduce a larger overhead.
Similar to the autoregressive mode, model weights need to be double buffered for $8$ and $16$ chips, whereas for $32$ and $64$ chips, on-chip memory is sufficient to hold all model weights, which results in reduced inference energy as can be seen from \figref{fig:plots_energy_latency} (b).

Overall, the results demonstrate the scalability of our partitioning scheme for Transformer-based models, especially for models dominated by off-chip transfer to higher-level of the memory hierarchy, such as the autoregressive TinyLlama model for which we achieve super-linear speedup for 8-32 chips and a quasi speedup for 64 chips.

\section{Conclusion}
\label{sec:conclusion}

In this paper, we presented a partitioning scheme tailored for deploying Transformer models on edge devices. With an approach inspired by tensor parallelism, this partition does not replicate any model weights across chips and only requires two synchronizations between chips, which allows the deployment of larger Transformer models at the extreme edge. We benchmark the partitioning scheme on the TinyLlama and MobileBERT models and demonstrated an above linear speedup of $26.1 \times$ for autoregressive and $9.9 \times$, for autoregressive and prompt TinyLlama mode, respectively, when using $8$ chips instead of a single chip system. For MobileBERT, we achieve a speedup of $4.7 \times$. To demonstrate the applicability to larger models, we showcase the scalability of our model. This work contributes to the active research field of deploying powerful Transformer-based models in highly resource-constraint devices. 

\section{Acknowledgement}
\label{sec:acknowledgement}

This work has received funding from the Swiss State Secretariat for Education, Research, and Innovation (SERI) under the SwissChips initiative. This work is funded in part by the dAIEDGE (\#101120726) and CONVOLVE (\#101070374) projects supported by the  EU Horizon Europe research and innovation program.


\bibliographystyle{IEEEtran}
\bibliography{ref}

\end{document}